\documentclass[twocolumn,showpacs,preprintnumbers,amsmath,amssymb,prb]{revtex4}

\usepackage{graphicx}
\usepackage{dcolumn}
\usepackage{bm}

\begin{document}

\title{Metal-insulator transition in PrRu$_4$P$_{12}$ and 
SmRu$_4$P$_{12}$ investigated by optical spectroscopy}

\author{M.~Matsunami}
 \altaffiliation[Present address: ]{RIKEN/SPring-8, 1-1-1 
Kouto, Mikazuki-cho, Sayo-gun, Hyogo 679-5148.}
 \email[E-mail: ]{matunami@spring8.or.jp}
\author{L.~Chen}
\author{M.~Takimoto}
\author{H.~Okamura}
\author{T.~Nanba}
\affiliation{Graduate School of Science and Technology, 
Kobe University, Kobe 657-8501, Japan.}

\author{C.~Sekine}
\author{I.~Shirotani}
\affiliation{Department of Electrical and Electronic Engineering, 
Muroran Institute of Technology, 27-1, Mizumoto, 
Muroran 050-8585, Japan.}

\date{\today}

\begin{abstract}
Electronic structures of the filled-skutterudite compounds PrRu$_4$P$_{12}$ 
and SmRu$_4$P$_{12}$, 
which undergo a metal-insulator transition (MIT) at $T_{\rm MI}$ = 60 K 
and 16 K, respectively, have been studied by means of optical spectroscopy. 
Their optical conductivity spectra develop an energy gap of $\sim$ 10~meV 
below $T_{\rm MI}$. 
The observed characteristics of the energy gap are qualitatively different 
from those of the Kondo semiconductors. 
In addition, optical phonon peaks in the spectra show anomalies upon the MIT, 
including broadening and shifts at $T_{\rm MI}$ and an appearance of new 
peaks below $T_{\rm MI}$. 
These results are discussed in terms of density waves or orbital ordering 
previously predicted for these compounds. 
\end{abstract}

                             
\maketitle
Ternary compounds $RM_4X_{12}$ ($R$ = rare earth elements; $M$ = Fe, Ru, Os; 
$X$ = P, As, Sb) with the filled-skutterudite structure 
(space group $Im\bar3$) exhibit a wide variety of physical properties. 
Among them, PrRu$_4$P$_{12}$ and SmRu$_4$P$_{12}$ are known to undergo a 
metal-insulator transition (MIT) at $T_{\rm MI}$ $\sim$ 60~K 
(Ref.~\onlinecite{Pr_base}) and 16~K,\cite{Sm_base} respectively. 
For PrRu$_4$P$_{12}$, the magnetic susceptibility shows no anomaly at 
$T_{\rm MI}$,\cite{Pr_base} and the valence of Pr is 3+ and independent of 
temperature.\cite{Pr_XANES} 
Thus, the MIT in PrRu$_4$P$_{12}$ is driven neither by a magnetic transition 
nor by a valence transition. 
Recently, evidence for superlattice formation below $T_{\rm MI}$ has been 
found by electron and X-ray diffraction 
experiments.\cite{Pr_electron-diff,Pr_X-diff1,Pr_X-diff2} 
Also, the band calculation study points out that the Fermi surfaces of 
$R$Ru$_4$P$_{12}$ should have a strong tendency for 
nesting in spite of the isotropic crystal structure.\cite{Pr_band-calc} 
In fact, the band calculation for PrRu$_4$P$_{12}$ 
assuming the displacement of P ions predicts an energy 
gap at the Fermi level ($E_{\rm F}$) \cite{Pr_band-calc_gap}. 
From these results, the MIT in PrRu$_4$P$_{12}$ has been considered to 
result from a charge density wave (CDW) transition caused by Fermi surface 
nesting.\cite{Pr_electron-diff,Pr_band-calc,Pr_band-calc_gap,Pr_specifc} 
In contrast, for SmRu$_4$P$_{12}$, the magnetic susceptibility clearly shows 
an anomaly near $T_{\rm MI}$.\cite{Sm_base} 
Moreover, recent works have revealed that the MIT involves two successive 
transitions at $\sim$ 14~K and $\sim$ 16~K.\cite{Sm_suc1,Sm_suc2} 
The two transitions and their magnetic-field dependencies appear 
qualitatively very similar to those of CeB$_6$, 
which shows an antiferro-quadrupolar (AFQ) ordering at the higher transition 
temperature $T_{\rm Q}$ and an antiferromagnetic ordering at the lower 
one.\cite{CeB6_AFQ} 
Therefore, it has been suggested that the MIT in SmRu$_4$P$_{12}$ should be 
related to an orbital ordering.

In this work, we have used optical reflectivity spectroscopy to further 
probe the microscopic electronic structures associated with the MIT in 
PrRu$_{4}$P$_{12}$ and SmRu$_{4}$P$_{12}$.\cite{pre} 
The optical technique is a powerful means for studying the electronic 
structures near $E_{\rm F}$, and has been applied to several skutterudite 
compounds including $R$Fe$_4$P$_{12}$,\cite{RFe4P12_opt} 
CeRu$_4$Sb$_{12}$ and YbFe$_4$Sb$_{12}$,\cite{CeRu4Sb12_opt} and 
CeOs$_4$Sb$_{12}$.\cite{CeOs4Sb12_opt} 
The obtained spectra of PrRu$_4$P$_{12}$ and SmRu$_4$P$_{12}$ clearly show a 
gap formation below $T_{\rm MI}$. 
In addition, anomalies of optical phonon peaks are observed around 
$T_{\rm MI}$. 
Comparing the present results with other reported physical properties, 
we analyze the evolution of electronic structures responsible for the MIT 
in PrRu$_{4}$P$_{12}$ and SmRu$_{4}$P$_{12}$.

The polycrystalline samples of PrRu$_4$P$_{12}$ (Ref.~\onlinecite{Pr_base}) 
and SmRu$_4$P$_{12}$ \cite{Sm_base} used had sizes of 
$\sim$ 3$\times$3$\times$1~mm$^3$, and their surfaces were mechanically 
polished. 
Temperature-dependent reflectivity spectra [$R(\omega)$] were measured under 
near-normal incidence, 
using a Fourier-transform interferometer and thermal sources at photon 
energies between 8~meV and 2~eV. 
A gold or silver film deposited $in~situ$ onto the sample surface was used 
as a reference of reflectivity.\cite{evapolation} 
Between 2 and 30~eV at room temperature, a grating spectrometer and 
synchrotron radiation source were used at the beam line BL7B of UVSOR, 
Institute for Molecular Science. 
The optical conductivity $\sigma(\omega)$ was obtained from Kramers-Kronig 
transformation applied to the measured $R(\omega)$.\cite{dressel} 
A Hagen-Rubens or a constant extrapolation was used below 8~meV, 
and a $\omega^{-4}$ extrapolation above 30~eV.\cite{dressel}

\begin{figure}[t]
\begin{center}
\includegraphics[width=0.4\textwidth]{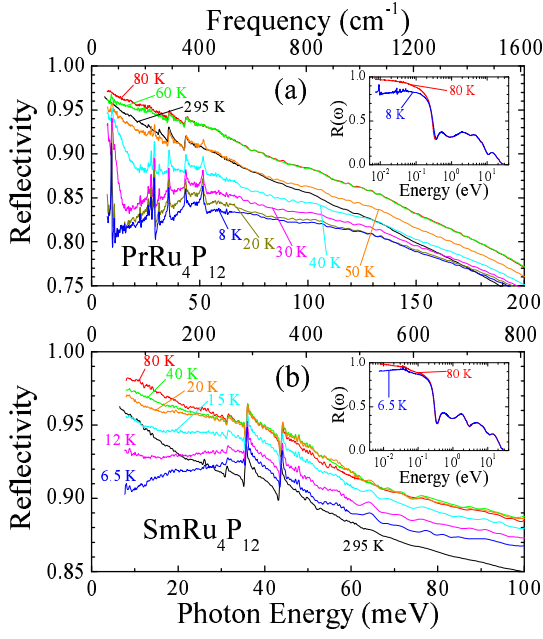}
\caption{
(Color online) Optical reflectivity spectra [$R(\omega)$] of 
PrRu$_4$P$_{12}$ (a) and SmRu$_4$P$_{12}$ (b) at different temperatures. 
Each inset shows $R(\omega)$ in a wider range of photon energies. 
} 
\end{center}
\end{figure}

\begin{figure}
\begin{center}
\includegraphics[width=0.45\textwidth]{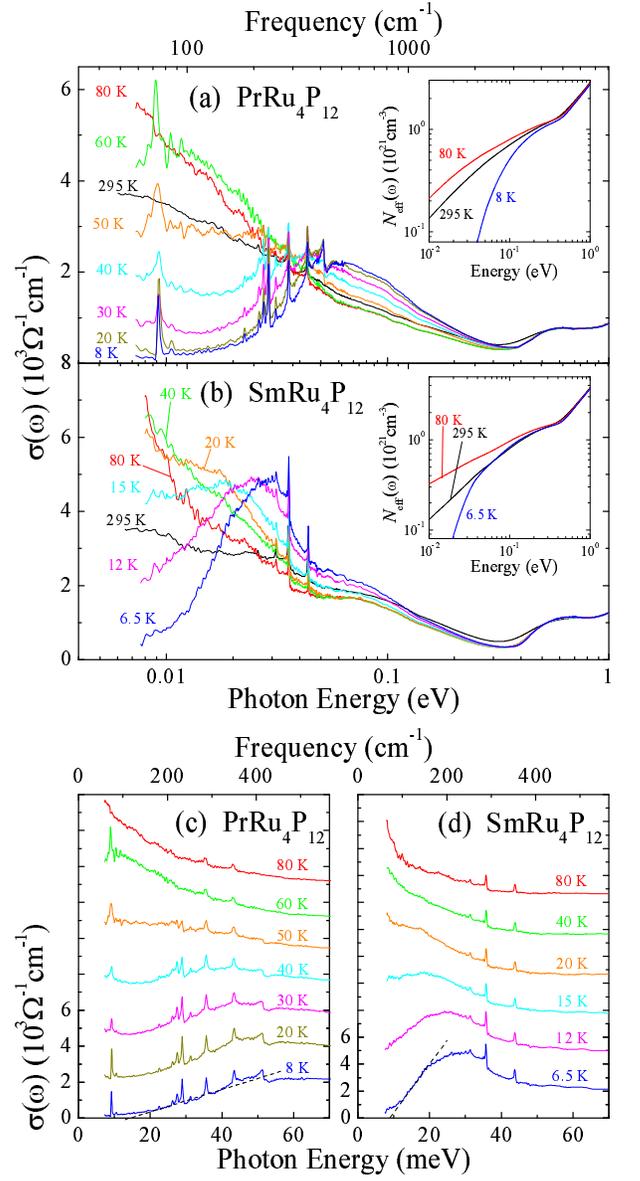}
\caption{
(Color online) Optical conductivity spectra $\sigma(\omega)$ of 
(a) PrRu$_4$P$_{12}$ and (b) SmRu$_4$P$_{12}$ at different temperatures.    
each inset shows the integrated spectral weight $N_{\rm eff}(\omega)$ 
(see text). 
The low-energy portion of the spectra in (a) and (b) are shown in (c) and (d), 
respectively. 
For clarity, 
each spectra are offset by $2 \times 10^3~\Omega^{-1} \rm{cm}^{-1}$ for (c) 
and $3 \times 10^3~\Omega^{-1} \rm{cm}^{-1}$ for (d). 
The dashed lines are fitted to the linearly rising portion of 
$\sigma(\omega)$. 
}
\end{center}
\end{figure}

Figure~1 shows the temperature ($T$) dependence of $R(\omega)$ for 
PrRu$_4$P$_{12}$ and SmRu$_4$P$_{12}$. 
The insets show $R(\omega)$ up to 30~eV. 
Between 295~K and 80~K, both compounds show typically metallic $R(\omega)$, 
with a plasma edge at $\sim$ 0.4~eV. 
The peaks above 0.4~eV are due to interband transitions. 
Below 80~K, in contrast, $R(\omega)$ decreases rapidly with decreasing $T$, 
indicating strong variations of the electronic structures near $E_{\rm F}$. 
In addition, sharp phonon peaks appear below 50~meV. 
The corresponding $\sigma(\omega)$ spectra are shown in Fig.~2. 
For both compounds, $\sigma(\omega)$ at 295 and 80~K are characterized by 
a Drude-type component due to free carriers, 
rising toward zero photon energy. 
Below 80~K, however, $\sigma(\omega)$ at low-energy region is suppressed, 
and an energy gap is progressively formed with decreasing $T$. 
By extrapolating the linearly varying portion of $\sigma(\omega)$ as shown 
in Fig.~2~(c) and (d), 
we estimate the magnitude of the energy gap to be $\sim$ 15~meV for 
PrRu$_4$P$_{12}$, and $\sim$ 10~meV for SmRu$_4$P$_{12}$. 
Associated with the gap formation, a broad peak grows around 60~meV for 
PrRu$_4$P$_{12}$, and around 30~meV for SmRu$_4$P$_{12}$. 
These peaks are due to optical excitations across the energy gap, 
and below we refer to them as the ``gap excitation peaks''. 
The effective carrier density, calculated as 
$N_{\rm eff}(\omega)$ = $(2m_0/\pi e^2) 
\int_{0}^{\omega}\sigma(\omega')d\omega'$,\cite{dressel} 
is plotted in the insets of Figs.~2 (a) and (b). 
They show that the optical sum rule is satisfied within a range of 
$\omega \sim$ 0.5~eV. 
This shows that the MIT is accompanied with 
the variation of electronic structures over $\sim$ 0.5~eV from $E_{\rm F}$, 
which is much larger than the gap magnitude.

\begin{figure}[t]
\begin{center}
\includegraphics[width=0.4\textwidth]{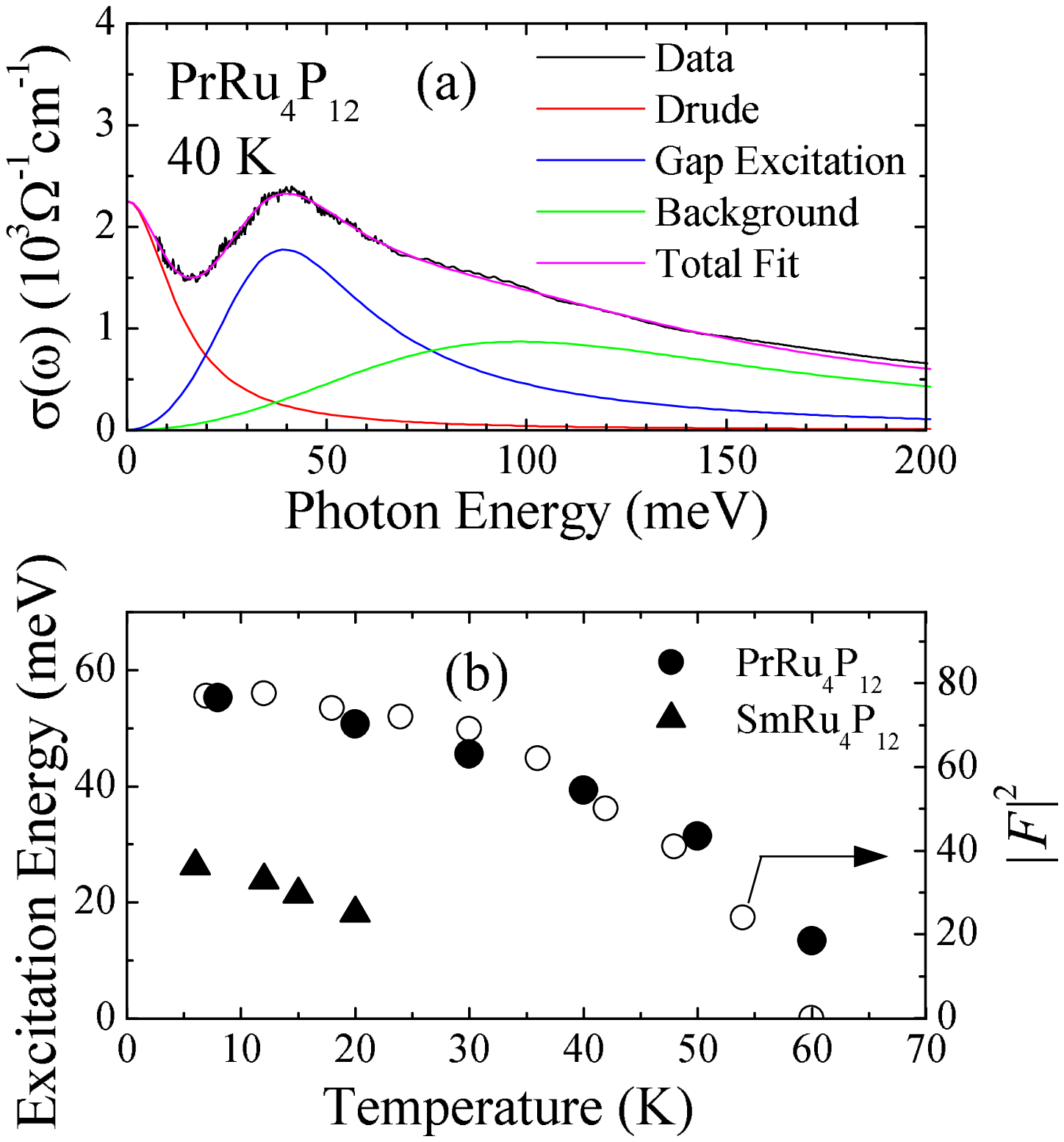}
\caption{
(Color online) (a) An example of the fitting for PrRu$_4$P$_{12}$, 
as discussed in the text. 
(b) The position of the gap excitation peak as a function of temperature. 
The superlattice diffraction intensity ($\mid$F$\mid$$^2$) of 
PrRu$_4$P$_{12}$ (open circles) is also shown\cite{Pr_X-diff1} for comparison. 
}
\end{center}
\end{figure}

As seen in Fig.~2, the energy gaps in these compounds are rather small, 
and there is a strong overlap with the Drude-like, 
free carrier contribution in $\sigma(\omega)$. 
Due to this, the estimate of the gap width described above cannot be applied 
at higher temperatures.  
Hence we identify the position of the gap excitation peak as the 
characteristic energy for the gap formation. 
To evaluate the peak position accurately, we have carried out spectral 
fittings, an example of which is shown in Fig.~3 (a) for PrRu$_4$P$_{12}$. 
We have used the Lorentz oscillator to fit the gap excitation peak, 
and the usual Drude function to fit the free-carrier component.\cite{dressel} 
To fit the background continuum, another Lorentz oscillator was placed at 
$\sim$ 100~meV for all $T$'s. 
The phonon peaks were also fitted as described later, 
and were subtracted out. 
Fig.~3~(b) shows the $T$-dependence of the peak position obtained by the 
fitting. 
It is clear that the peak energy increases with cooling for both 
PrRu$_4$P$_{12}$ and SmRu$_4$P$_{12}$. 
Also plotted in Fig.~3~(b) is the $T$-dependence of the intensity of a 
superlattice spot of PrRu$_4$P$_{12}$ investigated by 
X-ray diffraction.\cite{Pr_X-diff1} 
The $T$-evolution of the gap excitation peak closely follows that of 
the superlattice diffraction, hence the lattice deformation due to CDW. 
This result strongly suggests that the observed gap in PrRu$_4$P$_{12}$ is 
indeed related with the CDW formation. 
The observed characteristics of the gap for PrRu$_4$P$_{12}$ are in sharp 
contrast to those for the Kondo semiconductors such as 
Ce$_3$Bi$_4$Pt$_3$\cite{Ce343_opt} and YbB$_{12}$,\cite{YbB12_opt} 
where the gap width in $\sigma(\omega)$ is nearly unchanged with 
$T$.\cite{Ce343_opt,YbB12_opt}

For SmRu$_4$P$_{12}$, it is noteworthy that a decrease of $\sigma(\omega)$ is 
already seen at 20~K (below $\sim$ 15~meV), although a clear gap develops only 
below 16~K. 
The decrease of $\sigma(\omega)$ above $T_{\rm MI}$ probably indicates a 
precursor to the MIT, 
i.e., the density of states near $E_{\rm F}$ starts decreasing already above 
$T_{\rm MI}$. 
This is consistent with the result that the resistivity increases gradually 
with cooling below $\sim$ 50~K, 
before rapidly rising below $T_{\rm MI}$=16~K.\cite{Sm_base} 
In view of the prediction that the MIT in SmRu$_4$P$_{12}$ is related with 
an AFQ ordering,\cite{Sm_suc1,Sm_suc2} 
one possible origin for these results above $T_{\rm MI}$ is the fluctuation 
of the quadrupole moments above $T_{\rm Q}$ ($\simeq$~$T_{\rm MI}$): 
Although the long-range AFQ ordering can exist only below $T_{\rm Q}$, 
short-range ordering may exist even above $T_{\rm Q}$ with a strong 
fluctuation of quadrupole moments.\cite{AFQ_fluc} 
In fact, it has been reported that $T_{\rm Q}$ increases under magnetic 
field,\cite{Sm_suc1,Sm_suc2} which can be understood as resulting from a 
suppression of the fluctuations, 
similarly to the case of CeB$_6$.\cite{AFQ_fluc} 
Such strong fluctuation of the quadrupole moments may have reduced the 
density of states and $\sigma(\omega)$ above $T_{\rm MI}$.    
For example, for a perovskite oxide which forms an energy gap due to charge 
ordering, both a decrease in $\sigma(\omega)$ and an increase of 
resistivity with cooling have been observed above the ordering 
temperature.\cite{LSNO_opt} 
A fluctuation of charge order has been proposed as a possible origin for 
this case.

\begin{figure}[t]
\begin{center}
\includegraphics[width=0.45\textwidth]{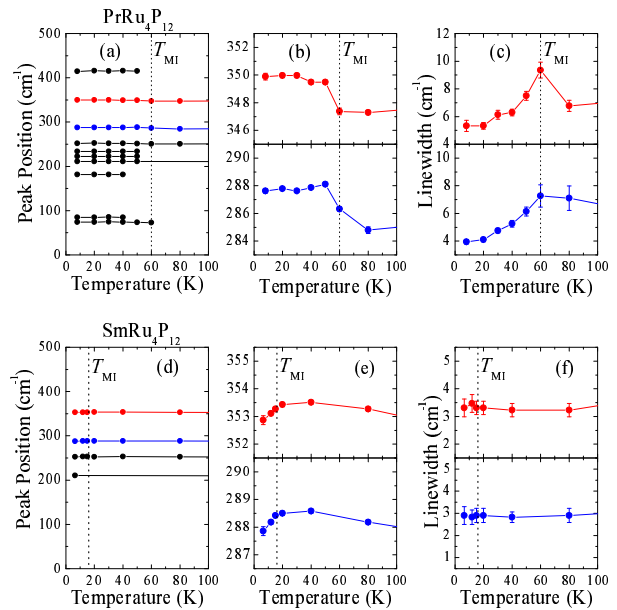}
\caption{(Color online) The peak frequencies and the linewidths 
of phonon peaks in $\sigma(\omega)$ as a function of temperauture 
for PrRu$_4$P$_{12}$ (a-c) and SmRu$_4$P$_{12}$ (d-f). }
\end{center}
\end{figure} 

Although the factor group analysis for the filled-skutterudite structure 
predicts 8 infrared-active phonon modes, 
the $\sigma(\omega)$ spectra of PrRu$_4$P$_{12}$ and SmRu$_4$P$_{12}$ at 
295~K show only 4 phonon peaks, similarly to other filled-skutterudite 
compounds.\cite{RFe4P12_opt,CeRu4Sb12_opt,CeOs4Sb12_opt} 
Below $T_{\rm MI}$ for PrRu$_4$P$_{12}$, however, several additional peaks 
appear in $\sigma(\omega)$, as shown in Fig.~2~(b) and plotted in Fig.~4~(a). 
The total number of observed phonon peaks below $T_{\rm MI}$ is larger 
than 8, which strongly suggests that the crystal structure of 
PrRu$_4$P$_{12}$ undergoes a symmetry lowering below $T_{\rm MI}$. 
To evaluate the detailed $T$ dependence of the phonons, 
we have fitted the phonon peaks using the Lorentz oscillator 
functions.\cite{YBCO_opt} 
The peak energies and the linewidths obtained from the fitting are plotted 
as a function of $T$ in Fig.~4. 
The phonon peaks near 285 and 347~cm$^{-1}$ for PrRu$_4$P$_{12}$, 
which are already present above $T_{\rm MI}$, show blue shifts of 
$\sim$ 3~cm$^{-1}$ and a narrowing of about 40~\% below $T_{\rm MI}$. 
These two phonon modes involve the vibrations of P.\cite{RFe4P12_opt} 
This result and the observation of more than 8 phonon peaks appear consistent 
with the slight displacement of P atoms below $T_{\rm MI}$ indicated by 
the superlattice diffraction 
data.\cite{Pr_electron-diff,Pr_X-diff1,Pr_X-diff2} 
Namely, displacement of P atoms may change the strength of bonding, 
leading to the peak shifts, and the associated symmetry lowering may result 
in more than 8 phonon peaks. 
The observed narrowing is likely to result from the rapid decrease of free 
carriers below $T_{\rm MI}$, 
since it should reduce the phonon damping due to carrier-phonon interaction. 
Clearly, the present results regarding the phonons are consistent with a 
CDW formation. 
It is noteworthy that the phonon peak characteristics of PrRu$_4$P$_{12}$ 
are qualitatively similar to those of 1$T$-TaS$_2$,\cite{TaS2_opt} 
which is a typical CDW compound. 
Namely, the infrared spectra of 1$T$-TaS$_2$ also show many additional 
phonon peaks, peak shifts, and peak narrowing below the CDW transition 
temperature.

The results of similar fitting for the phonons of SmRu$_4$P$_{12}$ are shown 
in Figs.4 (d)-(f). 
In contrast to PrRu$_4$P$_{12}$, 
no additional phonon peaks appear below $T_{\rm MI}$, 
and the peaks show red shifts below $T_{\rm MI}$ with almost no changes 
in the linewidth. 
The shifts are much smaller than those observed for PrRu$_4$P$_{12}$. 
Although a Fermi surface nesting is also predicted for SmRu$_4$P$_{12}$, 
the present result shows that the modulation 
in the charge density below $T_{\rm MI}$ is much weaker 
than that in PrRu$_4$P$_{12}$. 
This is consistent with the previous results that the ordering 
in SmRu$_4$P$_{12}$ should be of orbital or magnetic 
origin.\cite{Sm_suc1,Sm_suc2} 
In any case, the variations of the optical phonon peaks in $\sigma(\omega)$ 
upon the MIT are very different between 
PrRu$_4$P$_{12}$ and SmRu$_4$P$_{12}$. 
Remarkably, the elastic constants of these compounds, which are closely 
related to the acoustic phonons, have also shown very different variations 
around $T_{\rm MI}$.\cite{ultrasonic}

In conclusion, we have measured $\sigma(\omega)$ spectra of PrRu$_4$P$_{12}$ 
and SmRu$_4$P$_{12}$, to study the evolution of electronic structures 
upon the MIT. 
Their $\sigma(\omega)$ spectra have clearly shown the formation of an 
energy gap below $T_{\rm MI}$. 
For PrRu$_4$P$_{12}$, the $T$ evolution of the energy gap and the phonon 
peaks in $\sigma(\omega)$ are consistent with those associated with a 
CDW transition involving a symmetry lowering and a slight displacement of 
P atoms. 
For SmRu$_4$P$_{12}$, no clear sign of a density wave was observed in 
the evolution of $\sigma(\omega)$. 
The data suggest a decrease in the density of states even above 
$T_{\rm MI}$, which was discussed in terms of short-range orbital ordering 
preceding the MIT. 
The present results strongly suggest that the origin of the MIT is different 
between these compounds, similarly to the prediction by other experiments.

We thank Dr. A.~Irizawa, Dr. N. Ogita, Dr. C.~H.~Lee, Dr. Y.~Nakanishi, 
Prof. M.~Yoshizawa, and Prof. H.~Harima for helpful discussions. 
This work was partly supported by a Grant-in-Aid for Scientific Research 
Priority Area "Skutterudite" (Nos. 15072201 and 15072204) of the Ministry 
of Education, Culture, Sports, Science and Technology, Japan. 
A part of this work was performed as a Joint Studies Program of Institute 
for Molecular Science (2002).




\begin{thebibliography}{99}

\bibitem{Pr_base} 
C.~Sekine, T.~Uchiumi, I.~Shirotani, and T.~Yagi, 
Phys. Rev. Lett. {\bf 79}, 3218 (1997). 

\bibitem{Sm_base} 
C.~Sekine, T.~Uchiumi, I.~Shirotani, and T.~Yagi, 
Science and Technology of High Pressure, Universities Press, Hyderbad, 
India, 826 (2000). 

\bibitem{Pr_XANES}
C.~H.~Lee, H.~Oyanagi, C.~Sekine, I.~Shirotani, and M.~Ishii, 
Phys. Rev. B {\bf 60}, 13253 (1999). 

\bibitem{Pr_electron-diff}
C.~H.~Lee, H.~Matsuhira, A.~Yamamoto, T.~Ohta, H.~Takazawa, K.~Ueno, 
C.~Sekine, I.~Shirotani, and T.~Hirayama, 
J. Phys.: Condens. Matter {\bf 13}, L45 (2001). 

\bibitem{Pr_X-diff1}
L.~Hao, K.~Iwasa, K.~Kuwahara, M.~Kohgi, S.~R.~Saha, H.~Sugawara, Y.~Aoki, 
H.~Sato, C.~Sekine, C.~H.~Lee, and H.~Harima, 
J. Magn. Magn. Mater. {\bf 272-276}, Suppl. 1, E271 (2004). 

\bibitem{Pr_X-diff2}
C.~H.~Lee, H.~Matsuhata, H.~Yamaguchi, C.~Sekine, K.~Kihou, T.~Suzuki, 
T.~Noro, and I.~Shirotani, 
Phys. Rev. B {\bf 70}, 153105 (2004). 

\bibitem{Pr_band-calc} 
H.~Harima and K.~Takegahara, 
Physica B {\bf 312-313}, 843 (2002). 

\bibitem{Pr_band-calc_gap}
H.~Harima, K.~Takegahara, K.~Ueda, and S.~H.~Curnoe, 
Acta Physica Polonica B {\bf 34}, 1189 (2003). 

\bibitem{Pr_specifc}
K.~Matsuhira, Y.~Hinatsu, C.~Sekine, and I.~Shirotani, 
Physica B {\bf 312-313}, 829 (2002). 

\bibitem{Sm_suc1}
K.~Matsuhira, Y.~Hinatsu, C.~Sekine, T.~Togashi, H.~Maki, I.~Shirotani, 
H.~Kitazawa, T.~Takamasu, and G.~Kido, 
J. Phys. Soc. Jpn. {\bf 71} Suppl. 237 (2002). 

\bibitem{Sm_suc2}
C.~Sekine, T.~Uchiumi, K.~Matsuhira, P.~Haen, S.~D.~Brion, G.~Chouteau, 
H.~Suzuki, and H.~Kitazawa, 
Acta Physica Polonica B {\bf 34}, 983 (2003). 

\bibitem{CeB6_AFQ}
J.~M.~Effantin, J.~Rossat-Mignod, P.~Burlet, H.~Bartholin, S.~Kunii, and 
T.~Kasuya, 
J. Magn. Magn. Mater. {\bf 47~\verb|&|~48}, 145 (1985). 

\bibitem{pre}
A preliminary part of this work appeared in 
M.~Matsunami, L.~Chen, H.~Okamura, T.~Nanba, C.~Sekine, and I.~Shirotani, 
J. Magn. Magn. Mater. {\bf 272-276}, Suppl. 1, E39 (2004). 

\bibitem{RFe4P12_opt}
S.~V.~Dordevic, N.~R.~Dilley, E.~D.~Bauer, D.~N.~Basov, M.~B.~Maple, and 
L.~Degiorgi, 
Phys. Rev. B {\bf 60}, 11321 (1999). 

\bibitem{CeRu4Sb12_opt} 
S.~V.~Dordevic, D.~N.~Basov, N.~R.~Dilley, E.~D.~Bauer, and M.~B.~Maple, 
Phys. Rev. Lett. {\bf 86}, 684 (2001). 

\bibitem{CeOs4Sb12_opt} 
M.~Matsunami, H.~Okamura, T.~Nanba, H.~Sugawara, and H.~Sato, 
J. Phys. Soc. Jpn. {\bf 72}, 2722 (2003). 

\bibitem{evapolation} 
C. Homes, M. A. Reedyk, D. A. Crandels, and T. Timusk, 
Appl. Opt. {\bf 32}, 2976 (1993). 

\bibitem{dressel} 
M. Dressel and G. Gr{\"u}ner, 
{\it Electrodynamics of Solids} (Cambridge University Press, Cambridge, 2002). 

\bibitem{Ce343_opt} 
B. Bucher, Z. Schlesinger, P.C. Canfield, and Z. Fisk, 
Phys. Rev. Lett. {\bf 72}, 522 (1994).  

\bibitem{YbB12_opt} 
H. Okamura, S. Kimura, H. Shinozaki, T. Nanba, 
F. Iga, N. Shimizu, and T. Takabatake, 
Phys. Rev. B, {\bf 58}, R7496 (1998).   

\bibitem{AFQ_fluc}
N.~Fukushima and Y.~Kuramoto: 
J. Phys. Soc. Jpn. {\bf 67}, 2460 (1998). 

\bibitem{LSNO_opt}
T.~Katsufuji, T.~Tanabe, T.~Ishikawa, Y.~Fukuda, T.~Arima, and Y.~Tokura, 
Phys. Rev. B {\bf 54}, R14230 (1996).

\bibitem{YBCO_opt} 
The asymmetric line shapes of some of the peaks were treated 
using a fitting function discussed in 
J.~Sch{\"u}tzmann, S.~Tajima, S.~Miyamoto, Y.~Sato and R.~Hauff, 
Phys. Rev. B {\bf 52}, 13665 (1995). 

\bibitem{TaS2_opt}
L.~V.~Gasparov, K.~G.~Brown, A.~C.~Wint, D.~B.~Tanner, H.~Berger, 
G.~Margaritondo,  R.~Ga{\'a}l, and L.~Forr{\'o}: 
Phys. Rev. B {\bf 66}, 094301 (2002). 

\bibitem{ultrasonic}
M.~Yoshizawa, Y.~Nakanishi, T.~Kumagai, M.~Oikawa, 
C.~Sekine, and I~Shirotani, 
J. Phys. Soc. Jpn. {\bf 73}, 315 (2004). 

\end{thebibliography}
\end{document}